\documentclass[aps,showpacs,superscriptaddress,nofootinbib,floatfix]{revtex4-1}
\usepackage{graphicx}
\usepackage{epsf}
\def\slashchar#1{\setbox0=\hbox{$#1$}
   \dimen0=\wd0 \setbox1=\hbox{/} \dimen1=\wd1
   \ifdim\dimen0>\dimen1 \rlap{\hbox to \dimen0{\hfil/\hfil}} #1
   \else  \rlap{\hbox to \dimen1{\hfil$#1$\hfil}} / \fi}
\newcommand{\beas}{\begin{eqnarray*}}
\newcommand{\eeas}{\end{eqnarray*}}
\newcommand{\bea}{\begin{eqnarray}}
\newcommand{\eea}{\end{eqnarray}}

\begin{document}

\title{Single $\pi$ production in neutrino nucleus scattering}

\author{E.  \surname{Hern\'andez}} 
\affiliation{Departamento
de F\'\i sica Fundamental e IUFFyM,\\ Universidad de Salamanca, E-37008
Salamanca, Spain.}  
\author{J.  \surname{Nieves}}
\affiliation{ Instituto de F\'\i sica Corpuscular (IFIC), Centro Mixto
  CSIC-Universidad de Valencia, Institutos de Investigaci\'on de
  Paterna, Apartado 22085, E-46071 Valencia, Spain} 
\author{M. J. \surname{Vicente Vacas} }
\affiliation{Departamento de F\'\i sica Te\'orica e IFIC, Centro Mixto
  Universidad de Valencia-CSIC, Institutos de Investigaci\'on de
  Paterna, Apartado 22085, E-46071 Valencia, Spain}
  

\today

\begin{abstract}
We study $1\pi$ production in both charged and neutral current
neutrino nucleus scattering for neutrino energies below 2\,GeV. We use
a theoretical model for one pion production at the nucleon level that
we correct for medium effects. The results are incorporated into a
cascade program that apart from production also includes the pion
final state interaction inside the nucleus. Besides, in some specific
channels coherent $\pi$ production is also possible 
and we evaluate its contribution as well. Our results for total and differential cross
sections are compared with recent data from the MiniBooNE
Collaboration. The model provides an overall acceptable
 description of data, better for $NC$ than for $CC$ channels, although
 theory is systematically below data. Differential
 cross sections, folded with the full neutrino flux, show that most of
 the missing pions lie in the forward direction and at high energies. 
\end{abstract}

\maketitle

\section{Introduction}

A correct understanding of neutrino-nucleus interactions is crucial to
minimize systematic uncertainties in neutrino oscillation experiments
\cite{FernandezMartinez:2010dm}.  Most of the new generation of
neutrino experiments are exploring neutrino-nuclear scattering
processes at intermediate energies. Recently the MiniBooNE
Collaboration has published one pion production cross sections on
mineral oil by $\nu_\mu/\bar\nu_\mu$ neutrinos with energies below
2\,GeV. The data include neutral-current ($NC$) single $\pi^0$
production by $\nu_\mu$ and
$\bar\nu_\mu$~\cite{AguilarArevalo:2009ww}, as well as $\nu_\mu$
induced charged-current ($CC$) charged pion
production~\cite{AguilarArevalo:2010bm} and neutral pion
production~\cite{AguilarArevalo:2010xt}. These are the first pion
production cross sections to be measured since the old bubble chamber
experiments carried out at Argonne National Laboratory
(ANL)~\cite{Campbell:1973wg,Radecky:1981fn} and Brookhaven National
Laboratory (BNL)~\cite{Kitagaki:1986ct}. The latter were measured on
deuterium where nuclear effects are
small~\cite{AlvarezRuso:1998hi,Hernandez:2010bx}. The main
contribution to MiniBooNE data comes from $^{12}C$ and this poses an
extra problem to theoretical calculations as a direct test of any
fundamental production model is difficult since in-medium
modifications of the production mechanisms and final state interaction
(FSI) effects on the produced pions are important.

These new data show interesting deviations from the predictions of
present theoretical models~\cite{Morfin:2012kn}.  In
Ref.~\cite{Golan:2012wx}, the NuWro Monte Carlo event generator is
used to study $NC$ $\pi^0$ production data in nuclei.  A simple
theoretical model is used to describe the pion production on the
nucleon. No background terms are considered, and the coherent pion
production is calculated using the Rein-Sehgal
model~\cite{Rein:1982pf}, which is not appropriate at low
energies~\cite{Amaro:2008hd, Berger:2008xs,Hernandez:2009vm}.  A fair
agreement for neutrino induced reactions and a little worse agreement
for the antineutrino case are obtained.  $CC$ single pion production
off $^{12}C$ for neutrino energies up to 1\,GeV has been analyzed in
Ref.~\cite{Sobczyk.:2012zj}. The pion production theoretical model is
here more complete~\cite{Hernandez:2007qq} but nuclear effects have
been included only in a very simplified manner. Only total cross
sections were calculated and approximately agree with MiniBooNE data
for the $\pi^+$ channel and deviate more for the $\pi^0$ channel.  The
most comprehensive approach till now, with both a quite complete
microscopic description of the pion production on the
nucleon\footnote{The model~\cite{Leitner:2008ue} includes the weak
  excitation of several resonance contributions and their subsequent
  decay into $\pi N$. The vector couplings are taken directly over
  from the MAID analysis~\cite{MAID}, while the axial couplings are
  obtained from  partial conservation of the axial current (PCAC). For
  the necessary background terms, the vector part is again determined
  using the MAID analysis as a basis.  For the axial part (including
  the vector-axial interference) it was then assumed that it is
  proportional to the vector part. The proportionality constant is
  adjusted to the old bubble chamber ANL and BNL data, neglecting
  small deuteron effects.}  and the nuclear medium effects, can be
found in Ref.~\cite{Lalakulich:2012cj}. There, the Giessen
Boltzmann-Uehling-Uhlenbeck (GiBUU) model is used finding that total
cross sections measured by MiniBooNE are higher than theoretical ones
for neutrino energies above $0.8\sim0.9$\,GeV and obtaining also some
discrepancies in the differential cross sections. The study of
~\cite{Lalakulich:2012cj} is limited only to the incoherent part of
the $CC$ induced reaction, and no comparison with the MiniBooNE $NC$
data of Ref.~\cite{AguilarArevalo:2010bm} is performed. However, such
comparison was presented/discussed in NUFACT and NUINT conferences in
2009~\cite{Leitner:2009de,Leitner:2009ec}.

In this paper, we address the problem of $\nu(\bar{\nu})$-induced pion
production, both for $CC$ and $NC$ driven processes, using a more
sophisticated theoretical model, although it is restricted to
relatively low pion energies. We start from the one-pion production
model on nucleons from Refs.~\cite{Hernandez:2010bx} and
\cite{Hernandez:2007qq} which includes the $\Delta$ resonance
mechanisms, but also the background terms generated by the leading
order chiral Lagrangian. In order to extend the model to higher
energies above the $\Delta$ resonance region for which it was
originally developed, we add a new resonant contribution corresponding
to the $D_{13}(1520)$. According to Ref.~\cite{Leitner:2008ue}, this
resonance, besides the $\Delta(1232)$, is the only one playing a
significant role for neutrino energy below 2\,GeV.  Next, we have also
incorporated several nuclear medium corrections that directly affect
the production mechanisms. Apart from Pauli-blocking and Fermi motion
we take into account the important corrections that stem from $\Delta$
resonance properties modification inside the nuclear medium.  Finally,
the pion FSI is also relevant for the comparison with the experiment,
as a number of pions will be absorbed or re-scattered, possibly
changing their charge, in their way out of the nucleus. For the
inclusion of these effects, we shall follow Ref.~\cite{Salcedo:1987md}
where a simulation code for inclusive pion nucleus reactions was
developed.  In some specific channels coherent $\pi$ production in
$^{12}C$ is also possible and to evaluate its contribution, we use the
model of Ref.~\cite{Amaro:2008hd} but with the newer
nucleon-to-$\Delta$ form factors extracted in
Ref~\cite{Hernandez:2010bx}, where a simultaneous fit to ANL and BNL
data, accounting for deuteron effects as well, was carried out.
  
The paper in organized as follows: In Sec.~\ref{sec:piprod} we
introduce our model for $\pi$ production induced by $\nu(\bar{\nu})$
at the nucleon level, the relevant modifications for in-medium
calculations and briefly describe the implementation of the $\pi$ FSI.
A comparison with MiniBooNE data is shown in
Sec.~\ref{sec:results} and the main conclusions of this work are
collected in Sec.~\ref{sec:conclusions}. In Appendix~\ref{app:d13}, the details on the
$D_{13}$ contribution to pion production by neutrinos at the nucleon
level is included.

\section{Model for one pion production induced by neutrinos}
\label{sec:piprod}
\subsection{Pion production at the nucleon level}
\begin{figure}[tbh]
\centerline{\includegraphics[height=8cm]{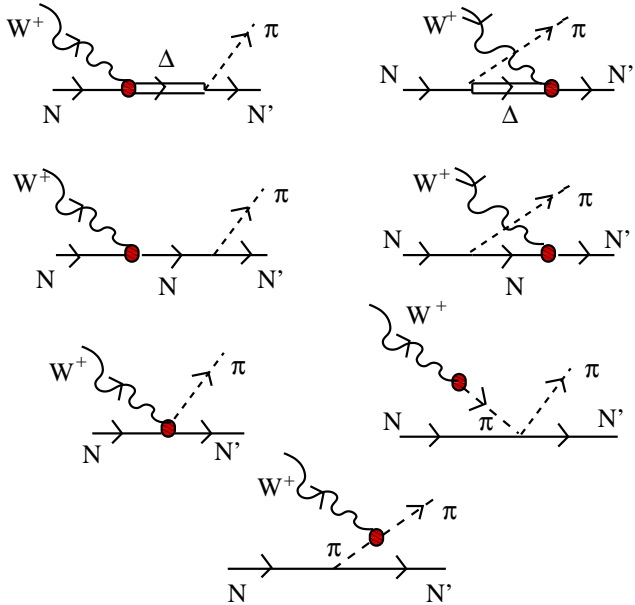}}
\caption{ Model for the $W^+N\to N^\prime\pi$
  reaction. We have direct and crossed
  $\Delta(1232)-$  and nucleon  pole terms,
  contact and pion pole contribution, and  the
  pion-in-flight term. We denote these
  contributions by: $\Delta P$, $C\Delta P$, $NP$, $CNP$, $CT$, $PP$ and
  $PF$, respectively.  }
  \label{fig:diagramas}
\end{figure}

The starting point  is the
 model of Ref.~\cite{Hernandez:2007qq} for one pion production on  the nucleon
  which is depicted diagrammatically  
 in Fig.~\ref{fig:diagramas}. 
It contains the dominant $\Delta$ resonance term and background terms
required by chiral symmetry. In total we have direct and crossed
$\Delta(1232)-$ (first row) and nucleon$-$ (second row) pole terms,
contact and pion pole contribution (third row) and finally the
pion-in-flight term. The background terms are the leading
contributions of a $SU(2)$ nonlinear $\sigma$ model. Those were
supplemented with well known form factors in a way that respected both
 conservation of the vector current (CVC) and PCAC hypotheses. Their
contribution is sizeable even at the $\Delta(1232)−$resonance 
peak and it turns out to be dominant near pion threshold.

 For the nucleon to $\Delta$ weak transition matrix element the form
 factor parametrization of Refs.~\cite{Llewellyn
   Smith:1971zm,Schreiner:1973ka} was taken\footnote{Note that the
   $C_5^A$ sign is quoted incorrectly in Ref.~\cite{Llewellyn
     Smith:1971zm} (see comment in Ref. ~\cite{Llewellyn
     Smith:1971zm,Schreiner:1973ka}).} in \cite{Hernandez:2007qq}. A
 total of four vector and four axial form factors are needed. The
 three vector form factors $ C_3^V(q^2),\ C_4^V(q^2), C_5^V(q^2)$ were
 determined from photo and electroproduction data whereas
 $C_6^V(q^2)=0$ from CVC. The vector form factors from
 Ref.~\cite{Lalakulich:2006sw} were used. For the axial part, Adler's
 model~\cite{Adler:1968tw} in which
 $C_3^A(q^2)=0,\ C_4^A(q^2)=-C_5^A(q^2)/4$, and PCAC, that requires
 $C_6^A(q^2)=-C_5^A(q^2)M^2/(m_\pi^2-q^2)$ with $M,\ m_\pi$ the
 nucleon and pion masses, were used. Thus, there is only one axial
 form factor to be determined, namely the dominant $C_5^A(q^2)$. For
 the latter, the parametrization of Ref.~\cite{Paschos:2003qr} was
 assumed in \cite{Hernandez:2007qq} and the unknown parameters
 $C_5^A(0)$ and $M_{A\Delta}$ were fitted to the flux averaged,
 $W<1.4$\,GeV, $\nu_\mu p\to\mu^-p\pi^+$\ $q^2$ differential cross
 section measured at ANL~\cite{Radecky:1981fn}. Here, $W$ stands for
 the final pion-nucleon invariant mass.
  
\begin{figure}
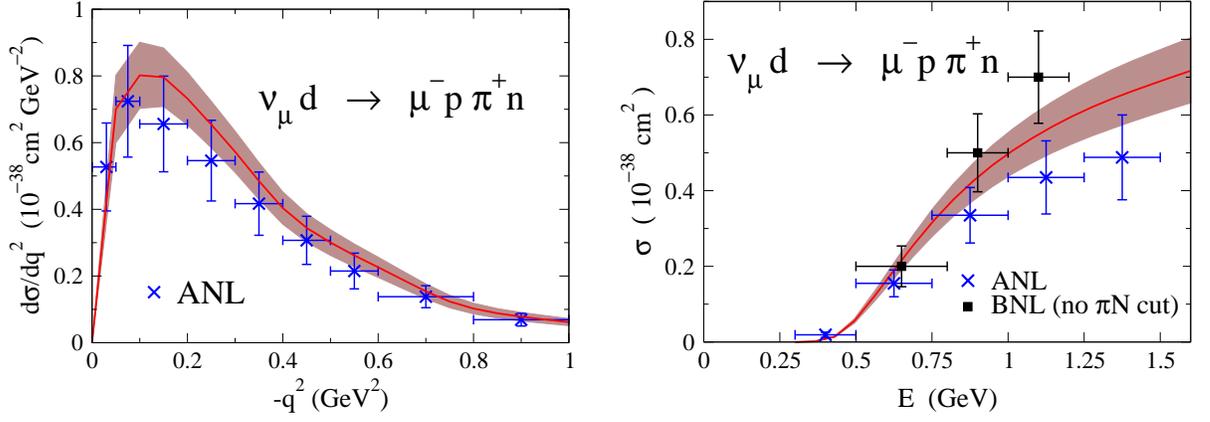

\center
\makebox[0pt]{\hspace{1cm}\includegraphics[height=5.5cm]{nuint10_1.eps}\hspace{.5cm}
\includegraphics[height=5.5cm]{nuint10_2.eps}}
\caption{Comparison of the theoretical model~\cite{Hernandez:2010bx}
  results (solid line) to ANL~\cite{Radecky:1981fn} and
  BNL~\cite{Kitagaki:1986ct} experimental data. Theoretical 68\%
  confidence level bands are also shown. Data include a systematic
  error (20\% for ANL and 10\% for BNL data) that has been added in
  quadratures to the statistical published errors. The theoretical
  results and ANL data include a $W<1.4$\,GeV cut in the final $\pi N$
  invariant mass. }
\label{fig:fit}
\end{figure}

While the results for different total and differential cross sections
were in good agreement with ANL
data~\cite{Radecky:1981fn,Campbell:1973wg}, the total cross sections
were smaller than the experimental data measured at
BNL~\cite{Kitagaki:1986ct}. BNL and ANL data seemed to be
incompatible. It was pointed out in Ref.~\cite{Graczyk:2009qm} that
this problem might originate from two factors: First, both ANL and BNL
data were measured on deuterium. Deuteron structure effects in the
$\nu_\mu d\to\mu^-\Delta^{++}n$ reaction were estimated in
Ref~\cite{AlvarezRuso:1998hi} to produce a reduction in the cross
section from $5-10\%$ and most analysis, including
Ref.~\cite{Hernandez:2007qq}, neglected that effect. Second, both
experiments suffered from neutrino flux uncertainties that in
Ref.~\cite{Graczyk:2009qm} were estimated to be $20\%$ for ANL and
10\% for BNL. Following the work of Ref.~\cite{Graczyk:2009qm}, a
combined fit of $C_5^A(q^2)$ to both ANL and BNL $p\pi^+$ data,  including
in both cases full deuteron effects and the flux normalization uncertainties,
was carried out in~\cite{Hernandez:2010bx}. In this latter fit, a simpler pure dipole
parameterization $C_5^A(q^2)=C_5^A(0)/(1-q^2/M_{A\Delta}^2)^2$ was
used obtaining $C_5^A(0)=1.00\pm0.11$ and $M_{A\Delta}=0.93\pm
0.07\,$GeV.  The results from that fit, compared to ANL and BNL data,
are shown in Fig.~\ref{fig:fit}.
The above mentioned parameterization for $C_5^A(q^2)$ is the one we are using 
in the present calculation.\\

For the present calculation, and in order to better compare with
MiniBooNE data, we need to extend the model up to $2\,$ GeV neutrino
energies, where higher mass resonances could play a role.  According
to Ref.~\cite{Leitner:2008ue}, and apart from the $\Delta$, only the
$D_{13}(1520)$ resonance gives a significant contribution in that
region.  We shall include in our model the two new contributions
depicted in Fig.~\ref{fig:d13}. All the details of the calculation of
these diagrams are given in Appendix~\ref{app:d13}.
\begin{figure}
\includegraphics[height=2.cm]{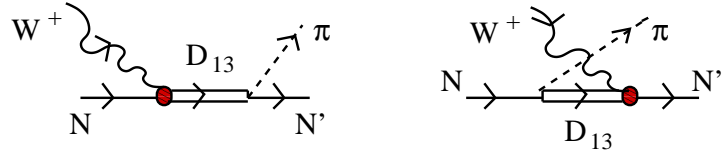}
\caption{$D_{13}(1520)$ contributions to 
 $W^+N\to N^\prime\pi$. We have direct ($DP$) and crossed ($CDP$) $D_{13}(1520)-$pole
 contributions. }
 \label{fig:d13}
\end{figure} 
We would like to remark that as the $D_{13}$ has isospin 1/2, it does not
contribute in the $p\pi^+$ channel and thus, it does not affect the previous
fit.\\

The differential 
neutrino-nucleon cross section with respect to the pion energy $E_\pi$ and 
the angle
between the pion momentum and   the neutrino beam $\theta_\pi$ 
is given for a $CC$ process by~\cite{Hernandez:2007qq}
\bea
\frac{d\sigma(\nu N\to l^-N'\pi)}{d\cos\theta_\pi dE_\pi}=2\pi\frac{G_F^2}{4\pi^2}
\frac{|\vec k_\pi|}{|\vec k|}
\frac1{4M}\frac1{(2\pi)^3}\int
d\Omega'dE'|\vec k'|\frac1{2E_{N'}}\,\delta(E_N+q^0-E_\pi-E_{N'})\,{\cal L}_{\mu\sigma}
(k,k'){\cal W}^{\mu\sigma}(p_N,q,k_\pi).\nonumber\\
\eea
$G_F$ is the Fermi decay constant. $k,\,k'$ and $k_\pi$ are the neutrino, final 
charged lepton and final pion 
four-momenta respectively. Besides, $q=k-k'$ is the four-momentum transferred, and 
$E_N=\sqrt{M^2+\vec p_N^{\,2}}$, 
$E_{N'}=\sqrt{M^2+(\vec p_N+\vec q-\vec k_\pi)^2}$ are the initial and 
final nucleon energies.
The lepton and hadronic tensors are given by
\bea
 {\cal L}_{\mu\sigma}(k,k')&=&k_\mu k'_\sigma+k_\sigma k'_\mu-k\cdot
k'g_{\mu\sigma}+i\epsilon_{\mu\sigma\alpha\beta}k^{\prime\alpha}k^\beta\\
{\cal W}^{\mu\sigma}(p_N,q,k_\pi)&=&\overline{\sum_{\rm spins}}\left\langle
N'\pi|j^\mu_{cc+}(0)|N\right\rangle\left\langle
N'\pi|j^\sigma_{cc+}(0)|N\right\rangle^*
\eea 
with $\epsilon_{0123}=+1$ and the metric $g_{\mu\nu}=(+,-,-,-)$, and
 where in the hadronic tensor  a sum over final  spins and an average
   over  initial ones is done.  The  $j^\mu_{cc+}(0)$ current contains all contributions 
 in Fig.~\ref{fig:diagramas}, see Refs.~\cite{Hernandez:2007qq,Hernandez:2010bx}
 for details,
 plus the new $D_{13}$ contributions of Fig.~\ref{fig:d13} that are given in
 Appendix~\ref{app:d13}. A similar expression is obtained for an $NC$ process in which
 we have a neutrino in the final state (see ~\cite{Hernandez:2007qq}).\\
 
\subsection{In-medium production}
 For incoherent production on a nucleus we have to sum the
 contribution to the cross section of all nucleons in the
 nucleus\footnote{For coherent production one should sum
   amplitudes~\cite{Amaro:2008hd}.}. Assuming the nucleus can be
 described by its density profile $\rho(r)=\rho_p(r)+\rho_n(r)$, and
 using the local density approximation, the initial differential cross section
 at the nucleus level for a pion production channel $N'\pi$, prior to any pion
 FSI, is then 
\bea 
\frac{d\sigma}{d\cos\theta_\pi dE_\pi}=\int
 d^3r\sum_{N=n,p}2\int \frac{d^3p_N}{(2\pi)^3}\ \theta(E_F^N(r)-E_N)
 \,\theta(E_N+q^0-E_\pi-E_F^{N'}(r)) \frac{d\sigma(\nu N\to
   l^-N'\pi)}{d\cos\theta_\pi dE_\pi}, 
\eea 
where $E_F^N(r)=\sqrt{M^2+(k_F^N(r))^2}$, being
 $k_F^N(r)=(3\pi^2\rho_N(r))^{1/3}$ the local Fermi momentum for
 nucleons of type $N$.  To compare with experiment, we have to
 convolute the above expression with the neutrino flux $\Phi(|\vec
 k|)$ 
\bea \frac{d\sigma}{\,d\cos\theta_\pi\, dE_\pi}&=&\int d|\vec
 k|\,\Phi(|\vec k|)\, 4\pi\int r^2\,dr \hspace{-.15cm}
 \sum_{N=n,p}\hspace{-.15cm}2\int
 \frac{d^3p_N}{(2\pi)^3}\ \theta(E_F^N(r)-E_N)
 \,\theta(E_N+q^0-E_\pi-E_F^{N'}(r)) \frac{d\sigma(\nu N\to
   l^-N'\pi)}{d\cos\theta_\pi dE_\pi}.\nonumber\\ 
\eea 
From there, we obtain 
\bea \frac{d\sigma}{d|\vec k|4\pi r^2\,dr\,d\cos\theta_\pi\,
   dE_\pi}=\Phi(|\vec k|)\,
 \hspace{-.15cm}
  \sum_{N=n,p}\hspace{-.15cm}2\int \frac{d^3p_N}{(2\pi)^3}\
\theta(E_F^N(r)-E_N) \,\theta(E_N+q^0-E_\pi-E_F^{N'}(r))
\frac{d\sigma(\nu N\to l^-N'\pi)}{d\cos\theta_\pi dE_\pi}.\nonumber\\
\eea
Apart from modifications discussed in what follows, the above differential
cross section is used in
our simulation code to generate, in a given point inside the nucleus and by
neutrinos of a given energy, pions with a certain charge, energy and momentum 
direction.

Defining $P=q-k_\pi$ (the four momentum transferred to the nucleus)
and writing $d^3p_N=d\cos\vartheta_N\, d\phi_N\, |\vec p_N| E _N
dE_N$, where the angles are referred to a system in which the $Z$ axis
is along $\vec P$, we can integrate in the $\vartheta_N$ variable
using the energy delta function present in $\frac{d\sigma(\nu N\to
  l^-N'\pi)}{d\cos\theta_\pi dE_\pi}$. The final result is \bea
\frac{d\sigma}{d|\vec k|4\pi r^2\,dr\,d\cos\theta_\pi\, dE_\pi}
&=&\Phi(|\vec k|)\, \int d\Omega'dE'|\vec k'| \bigg\{
\sum_{N=n,p}\hspace{-.15cm}\frac{G_F^2}{512\pi^7} \frac{|\vec
  k_\pi|}{|\vec P\,|\ |\vec k|} \theta(E_F^N(r)-{\cal E})
\,\theta(-P^2)\,\theta(P^0)\,{\cal
  L}_{\mu\sigma}(k,k')\nonumber\\ &&\int_0^{2\pi} d\phi_N \int_{\cal
  E}^{E_F^N(r)}dE_N {\cal
  W}^{\mu\sigma}(p_N,q,k_\pi)\bigg|_{\cos\vartheta_N=\cos\vartheta_N^0}\bigg\},
\label{eq:gen0}
\eea
where
\begin{equation}
\cos\vartheta^0_N=\frac{P^2+2E_NP^0}{2|\vec p_N||\vec P|},\ 
{\cal E}'=\frac{-P^0+|\vec P|\sqrt{1-4M^2/P^2}}2,\ {\cal
E}=\max\{M,E_F^{N'}-P^0,{\cal E}'\} \label{eq:appx1}.
\end{equation}
To speed up the computational time, we approximate the last two
integrals in Eq.(\ref{eq:gen0}) by 
\bea
\int_0^{2\pi} d\phi_N \int_{\cal E}^{E_F^N(r)}dE_N
{\cal W}^{\mu\sigma}(p_N,q,k_\pi)\bigg|_{\cos\vartheta_N=\cos\vartheta_N^0}
\approx 2\pi(E_F^N(r)-{\cal E}){\cal W}^{\mu\sigma}(\tilde
p_N,q,k_\pi)\bigg|_{\cos\vartheta_N=\cos\vartheta_N^0} \label{eq:approx}
\eea 
where $\tilde p_N$ is evaluated at the value $\tilde E_N=(E_F^N(r)+{\cal E})/2$,
(middle of the integration interval), with the corresponding
$\cos\tilde\vartheta_N^0$ $\big($that deduced from Eq.~(\ref{eq:appx1})
using $E_N=\tilde E_N$ and $|\vec p_N|= \sqrt{\tilde E_N^2-M^2}$ $\big
)$, and $\tilde \phi_N$ is set
to zero. Similar approximations were done, and shown to be sufficiently accurate, in
 Refs. \cite{Carrasco:1989vq,Carrasco:1991mb, Gil:1997bm, Gil:1997jg}
to study  total inclusive  and total inclusive pion production 
in photon and electron nuclear reactions.  In the study in
Ref.~\cite{Nieves:2011pp} of total inclusive neutrino induced cross
section this kind of simplification was also used. We have checked
that the approximation of Eq.~(\ref{eq:approx})  induces uncertainties
at most of 5\%, independently of $\tilde \phi_N$.  Other choices to 
fix $\tilde \phi_N$ produce small variations of the order of 1-2\%.
With this approximation, we find
\bea
 \frac{d\sigma}{d|\vec k|4\pi  r^2\,dr\,d\cos\theta_\pi\, dE_\pi}
 &\approx&\Phi(|\vec k|)\,
\int
d\Omega'dE'|\vec k'| \bigg\{
  \sum_{N=n,p}\hspace{-.15cm}\frac{G_F^2}{256\pi^6}
\frac{|\vec k_\pi|}{|\vec P\,|\ |\vec k|}\,(E_F^N(r)-{\cal E})\,
\theta(E_F^N(r)-{\cal E}) \,\theta(-P^2)\,\theta(P^0)\nonumber\\
&&\hspace{3.5cm}\times{\cal L}_{\mu\sigma}(k,k')
{\cal W}^{\mu\sigma}(\tilde p_N,q,k_\pi)\bigg\}.
\label{eq:gen1}
\eea
Eq.(\ref{eq:gen1})  includes explicitly Fermi motion of the initial nucleon
and Pauli blocking of the final
nucleon but there are other important in-medium corrections, that we discuss  in the
following, that have to be included. The above expression is equivalent
to Eqs. (2) and (25) of Ref.~\cite{Nieves:2011pp}, where among others,
the 1p1h1$\pi$
excitations contribution to the total inclusive neutrino-nucleus cross
sections was evaluated.

\subsection{In medium corrections to pion production and final state interaction}
Given the dominant role played by the $\Delta P$ contribution and since 
$\Delta$ properties are strongly modified in the nuclear 
medium~\cite{Hirata:1978wp,Oset:1987re,Nieves:1993ev,Gil:1997bm,Benhar:2005dj,AlvarezRuso:2007tt,
Amaro:2008hd,AguilarArevalo:2010zc,Hernandez:2010jf} a more proper treatment of
the $\Delta$ contribution is needed. Here, we follow Ref.~\cite{Gil:1997bm}
and
modify the  $\Delta$ propagator in the $\Delta P$ term as
\beas
\frac1{p_\Delta^2-M_\Delta^2+iM_\Delta\Gamma_\Delta}\to\frac1{\sqrt{s}+M_\Delta}
\frac1{\sqrt{s}-M_\Delta+i(\Gamma^{\rm Pauli}_\Delta/2-{\rm Im}\Sigma_\Delta)},
\eeas
with $s=p_\Delta^2$, $\Gamma^{\rm Pauli}_\Delta$ the free $\Delta$ width
corrected by Pauli blocking of the final nucleon, for which we take 
the expression in Eq.(15) of
 Ref.~\cite{Nieves:1991ye}, and ${\rm Im}\Sigma_\Delta$
the imaginary part of the $\Delta$ self-energy in the medium. For the mass we shall
keep  its free value. While there are some corrections to the mass
coming both 
from the real part of
the self-energy and RPA sums, together they induce changes smaller than
the precision in the present experiments and the uncertainties due to our
limited knowledge of the nucleon to $\Delta$ transition form factor
$C_5^A(q^2)$, see discussion in Sec. II.E of Ref.~\cite{Nieves:2011pp}.

The evaluation of
$\Sigma_\Delta$ is done in Ref.~\cite{Oset:1987re} where the imaginary part is
parametrized as
\beas
-{\rm Im}\Sigma_\Delta=C_Q\left(\frac{\rho}{\rho_0}\right)^\alpha
+C_{A_2}\left(\frac{\rho}{\rho_0}\right)^\beta+
C_{A_3}\left(\frac{\rho}{\rho_0}\right)^\gamma,
\eeas
with $\rho_0=0.17\,$fm$^{-3}$. The $C_Q,\alpha$, $C_{A_2},\beta$ and 
$C_{A_3},\gamma$ coefficients can be found in Eq.(4.5)
 and Table 2 of Ref.~\cite{Oset:1987re}. They are parametrized as a function of
 the kinetic energy of a pion that would excite  a $\Delta$ of the corresponding
 invariant mass and are valid in the range $85\, {\rm MeV} < T_\pi < 315 \,{\rm
 MeV}$. Below 85\,MeV the contributions from $C_Q$ and $C_{A_3}$
are rather small and we take them from Ref.~\cite{Nieves:1991ye}, where the model 
was extended to low energies. The term with $C_{A_2}$ shows a very mild
energy dependence and we still use the parameterization from Ref.~\cite{Oset:1987re}
 even at low energies. For $T_\pi$ above 315 MeV, we have kept these
self-energy terms constant and equal to their values at the bound. 
The uncertainties in these pieces are not very relevant there because
the  $\Delta\to N\pi$ decay becomes very large and  dominant.

The terms in $C_{A_2}$ and $C_{A_3}$ are related to the two-body absorption
$WNN\to NN$ and three-body absorption $WNNN\to NNN$ channels respectively. On the other
hand the $C_Q$ term gives rise to a new $WN\to N\pi$ contribution   inside
the nuclear medium and thus it has to be taken into account beyond its role
in modifying 
the $\Delta$ propagator. This new contribution has to be added incoherently and
we implement it in a approximate way by taking as   amplitude square for 
this process the amplitude square of the
 $\Delta P$ contribution multiplied by
 \bea
 \frac{C_Q(\rho/\rho_0)^\alpha}{\Gamma^{\rm Free}_\Delta/2}\label{eq:cq}.
 \eea
 Our final model for production is then given in Eq.(\ref{eq:gen1}) with the
 modifications in the hadronic tensor just mentioned.\\

Once the pion is produced inside the nucleus, it starts propagating
and it suffers interactions with the medium.  To evaluate them we
follow Ref.~\cite{Salcedo:1987md}, where a computer simulation code
was developed to describe inclusive pion nucleus reactions
(quasielastic, single charge exchange, double charge exchange and
absorption). We take into account $P$- and $S$-wave pion absorption,
and $P$-wave quasielastic scattering on a single nucleon. The $P$-
wave interaction is mediated by the $\Delta$ resonance excitation
where the different contributions to the imaginary part of its
self-energy give rise to pion two- and three-nucleon absorption and
quasielastic processes. After a quasielastic interaction the pions
change direction and may change charge. The intrinsic probabilities
for each of the above mentioned reactions were evaluated
microscopically as a function of the density and we use the local
density approximation to evaluate them in finite nuclei. In between
collisions the pions are treated as classical particles and in the
present calculation we shall assume they propagate in straight lines.
All details of the simulation can be found in
Ref.~\cite{Salcedo:1987md}.  We should remark that this approach to
$\pi$ FSI has been extensively and successfully used in the study of
many processes such as hypernuclear decays~\cite{Oset:1989ey}, $\pi$
absorption~\cite{VicenteVacas:1993bk},
photon~\cite{Carrasco:1991mb,Carrasco:1992mg} and
electron~\cite{Gil:1997jg} induced nuclear processes (like $\pi$
production) or muon capture~\cite{Chiang:1989nh}.  Furthermore, it has
been extensively used in their analysis by the successive Kamiokande
collaborations. See,
e.g. Refs.~\cite{Ashie:2005ik,Ahn:2006zza,Abe:2011sj}.

\subsection{Coherent production}
In some of the channels there is a contribution from coherent $\pi$
production. We evaluate it with the model of Ref.~\cite{Amaro:2008hd},
using the $C_5^A(q^2)$ form factor obtained in
Ref~\cite{Hernandez:2010bx}. Cross sections for the T2K and MiniBooNE
fluxes obtained with this new form factor were given in
\cite{Hernandez:2010jf}.  The model for coherent pion production
is based on the microscopic model for pion production off the nucleon
of Ref.~\cite{Hernandez:2007qq} that as already mentioned, besides the
dominant $\Delta$ pole contribution, takes into account the effect of
background terms required by chiral symmetry. Our coherent production
model does not incorporate the $D_{13}$ contribution that we expect to
produce only a small correction. 

The main nuclear
effects, namely, medium corrections on the $\Delta$ propagator and the
final pion distortion, are included. As found in similar
calculations~\cite{AlvarezRuso:2007tt, Singh:2006bm, AlvarezRuso:2007it}, 
the modification of the $\Delta$ self-energy inside the nuclear medium
strongly reduces the cross section, while the final pion distortion
mainly shifts the peak position to lower pion energies.

We should stress that  the model of Ref.~\cite{Amaro:2008hd} 
is more reliable than the Rein-Sehgal approach ~\cite{Rein:1982pf}
 for the energies of interest in this work. In
particular, it greatly improves on the description of angular
distributions of the outgoing pion with respect to the direction of
the incoming neutrino~\cite{Amaro:2008hd,Hernandez:2009vm}.
\section{Results for pion production in neutrino nucleus scattering and
comparison with MiniBooNE data}
\label{sec:results}
In this section we compare our predictions with data recently obtained by the
MiniBooNE Collaboration for $\nu_\mu$ induced $CC$ 
~\cite{AguilarArevalo:2010bm,AguilarArevalo:2010xt} and 
$\nu_\mu/\bar\nu_\mu$ induced $NC$~\cite{AguilarArevalo:2009ww} pion production in mineral oil ($CH_2$). 
\subsection{$CC$ production}
\begin{figure}[tbh]
 \center
\includegraphics[height=7cm]{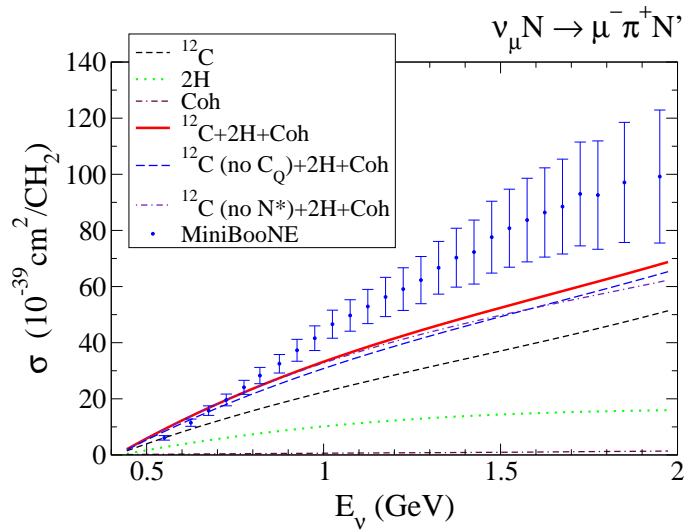}
\caption{ $1\pi^+$ total production cross section for $\nu_\mu$ $CC$
  interaction in mineral oil.  Short-dashed line: $^{12}C$
  contribution. Dotted line: $H_2$ contribution. Double-dashed dotted
  line: Coherent contribution (see main text).  Solid line: Total
  contribution. Long-dashed line: Same as solid line but without the
  $C_Q$ contribution of Eq.~(\ref{eq:cq}). Dashed-dotted line: Same as
  solid line but without the contribution from the $D_{13}$ resonance
  contribution. Data taken from Ref.~\cite{AguilarArevalo:2010bm}. }
  \label{fig:totalccpip}
\end{figure}

We start by showing results for the total unfolded cross sections for charged
current one-pion production. In Fig.~\ref{fig:totalccpip}, we compare
our results with the data by the MiniBooNE Collaboration for a final
$\pi^+$. We take into account the contribution on $^{12}C$ and that on
the two hydrogens. There is also a small coherent contribution on
$^{12}C$ that we have evaluated as described above.  Our total result
agrees well with data at low neutrino energies, but is below data for
neutrino energies above $0.9$\,GeV.

\begin{figure}[tbh]
 \center
\includegraphics[height=7cm]{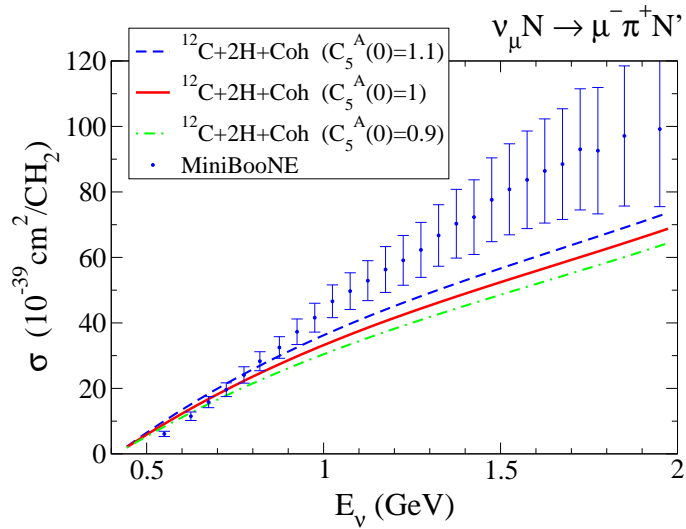}
\caption{ $1\pi^+$ total production cross section 
for $\nu_\mu$ $CC$ interaction
in mineral oil. Solid line: Our full model with $C_5^A(0)=1$. Dashed line: Full model with 
$C_5^A(0)=1.1$. Dashed-dotted line: Full model with 
$C_5^A(0)=0.9$.}
  \label{fig:totalccpipc5a}
\end{figure}

The $D_{13}$ contribution is only noticeable above $E_\nu=1.2$\,GeV.
At around $E_\nu=2$\,GeV, it makes some 8\% of the total
contribution. It will play a minor role for observables that are
convoluted over the MiniBooNE neutrino flux as the latter peaks at
around 0.6\,GeV. On the other hand, in the whole energy range shown,
the $C_Q$ term (Eq.~(\ref{eq:cq})) produces changes always smaller
than 10\%.  In Fig.~\ref{fig:totalccpipc5a}, we show the effect of varying
$C_5^A(0)$ within the uncertainties in its determination, using BNL
and ANL data, in Ref.\cite{Hernandez:2010bx}. We find effects at the
10\% level. The highest value seems to be favored by the MiniBooNE
data. We should remind here that if flux uncertainties were ignored,
BNL and ANL data would not be compatible~\cite{Graczyk:2009qm}.  Taken
at face value, the figure suggests that MiniBooNE data would support
BNL results. This would also imply larger values of $C_5^A(0)$ closer
to the PCAC prediction, as suggested in
Ref.~\cite{Lalakulich:2012cj}. However, one should be cautious given
the number of uncertainties in both experiment and theoretical models.

\begin{figure}[h!!]
 \center
\includegraphics[height=7cm]{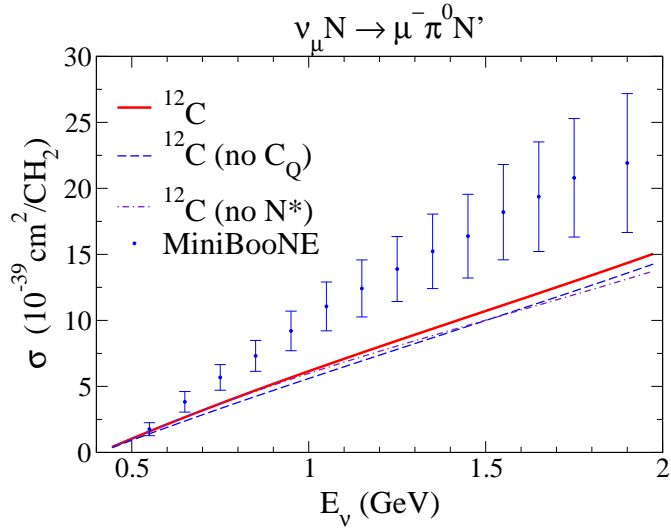}
\caption{ $1\pi^0$ total production cross section for $\nu_\mu$ $CC$
interaction
 in mineral oil. In this case there is no contribution from the $H_2$ and there
 is no coherent contribution. Captions as in Fig.~\ref{fig:totalccpip}. Data
 from Ref.~\cite{AguilarArevalo:2010xt}. }
  \label{fig:totalccpi0}
\end{figure}

 Similar results, consistently below data, are obtained for the case
 of CC $\pi^0$ production, see Fig.~\ref{fig:totalccpi0}. As in the
 previous case, the role of the $D_{13}$ resonance or the $C_Q$
 corrections are small.

\vspace{.1cm}

\begin{figure}[h!!]
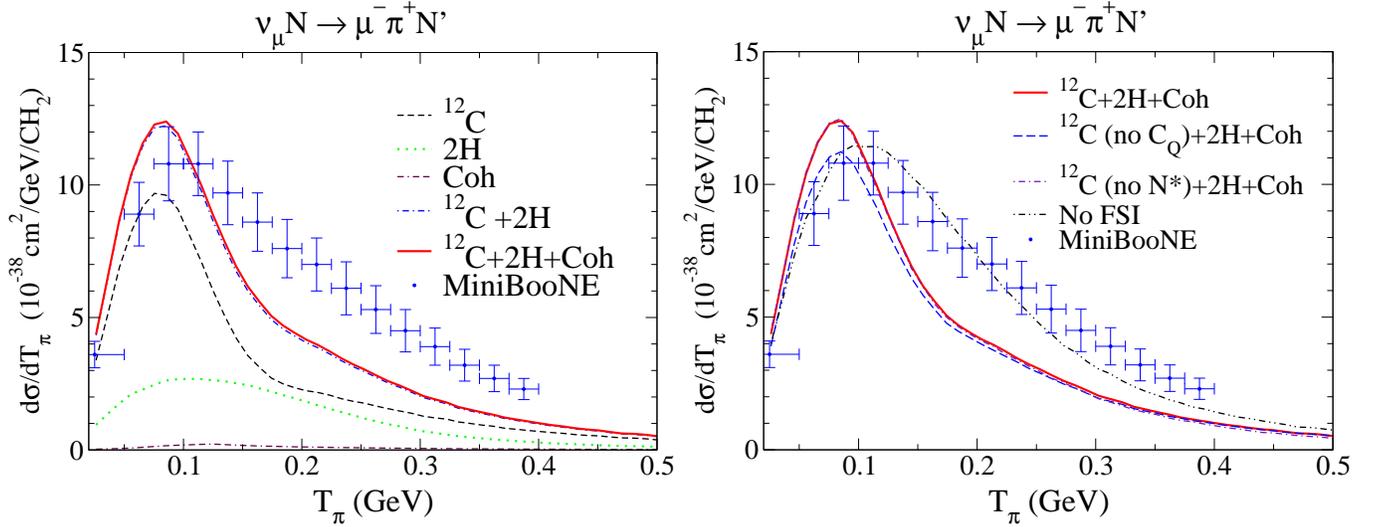

 \center
\includegraphics[height=7cm]{2ccpipKin.eps}
\includegraphics[height=7cm]{2ccpipKinV2.eps}
\caption{ Flux-folded differential $\frac{d\sigma}{dT_\pi}$ 
cross section for $CC$ $1\pi^+$ production by $\nu_\mu$ in mineral
oil. Captions as in Fig.~\ref{fig:totalccpip}, but in addition we
also display  results neglecting FSI (double-dotted dashed line in the
right panel) and coherent (short-dashed dotted line in the left panel) contributions. Data
 from Ref.~\cite{AguilarArevalo:2010bm}. Note that the coherent (left
 panel) and $D_{13}$ resonance (right panel) contributions are very
 small. The curves obtained when these effects are neglected can
 hardly be distinguished from  the red solid line that stands for the
 total contribution. }
  \label{fig:cctpi}
\end{figure}

In Fig.~\ref{fig:cctpi}, we compare the differential
$\frac{d\sigma}{dT_\pi}$ cross section for $CC$ $1\pi^+$ production by
$\nu_\mu$, calculated with the MiniBooNE flux from
Ref.~\cite{AguilarArevalo:2010bm}. In the left panel, we show the
different contributions to the full model, including the coherent
part. The model predicts less high energy pions than the experiment
for $T_\pi$ above 0.15\,GeV.  The combined effect of quasielastic
scattering and pion absorption through $\Delta$ excitation depletes
the $0.15\sim 0.4$\,GeV region. As a result the strength moves down to
0.8\,GeV. The shape of the calculated cross section is dominated by
the $^{12}C$ contribution and both the coherent part and the hydrogen
contribution peak at higher energies but, they are too small to
compensate for the missing high energy pions.  On the right panel we
also show the effect of not including the $C_Q$ correction of Eq.~(\ref{eq:cq}). We can
see, that it little affects the cross section and/or the shape of the
pion kinetic energy differential distribution.

From the above discussion we might expect a  better agreement with
data when FSI effects are neglected. Indeed, as can be seen on the right panel of
Fig.~\ref{fig:cctpi}, the experimental shape could be reproduced by
artificially removing the FSI of the pion.  Of course, this has little
sense as FSI is really there. Removing FSI completely
would lead to total cross section values that are too high at low
neutrino energies, where the theoretical model is more reliable. FSI
effects could be reduced by considering the so called ``formation
zone''~\cite{Golan:2012wx}, that among other effects includes the
propagation of the $\Delta$ before decaying into a $\pi N$
pair.  Of course, the ``formation
zone'' could be adjusted to reproduce data. These kind of modifications of the FSI
could be difficult to justify, they might be in conflict with much other
phenomenology and may somehow serve to hide our ignorance on the relevant dynamics. Even
when they could help reproducing some observable, if we lack a correct
understanding of the physical mechanisms responsible for them, they
might lead to wrong predictions for other observables sensible to
other kinematics, dynamical mechanisms or nuclear corrections.

\begin{figure}[tbh]
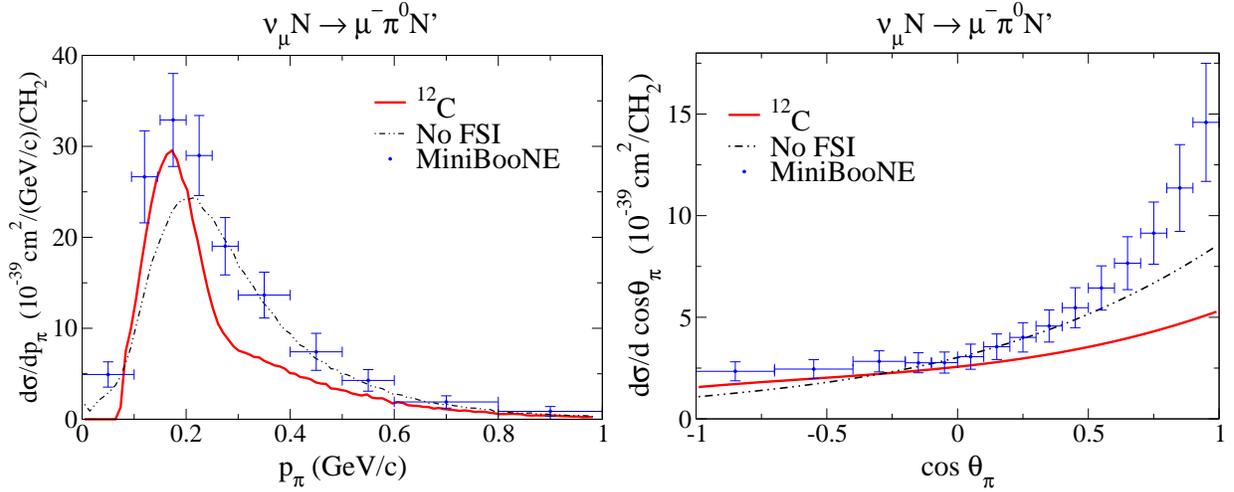

 \center
\includegraphics[height=6.5cm]{1ccpi0mom.eps}
\includegraphics[height=6.5cm]{1ccpi0cos.eps}
\caption{ Flux-folded differential $\frac{d\sigma}{dp_\pi}$ (left panel) and
$\frac{d\sigma}{d\cos\theta_\pi}$ (right panel)
cross section for $CC$ $1\pi^0$ production by $\nu_\mu$ in mineral
oil. Captions as in Fig.~\ref{fig:cctpi}. Data
 from Ref.~\cite{AguilarArevalo:2010xt}. }
  \label{fig:ccpi0}
\end{figure}

In Fig.\ref{fig:ccpi0}, we show the flux-folded differential
$\frac{d\sigma}{dp_\pi}$ and $\frac{d\sigma}{d\cos\theta_\pi}$ cross
sections for $CC$ $1\pi^0$ production by $\nu_\mu$ that we compare to
data. For that we use the neutrino flux reported in
Ref.~\cite{AguilarArevalo:2010xt}, that extends from 2\,GeV down to
0.5\,GeV neutrino energies. In this case, there is neither contribution
from the hydrogen nuclei nor from coherent production.  Once more, the
FSI effects are clearly visible in both distributions and their
artificial exclusion leads to a better description of the high
momentum tail of the $\frac{d\sigma}{dp_\pi}$ distribution.  Because
of the FSI some pions are absorbed, but other ones are scattered and
loose to nucleons part of their energy.  FSI is essential to fill the
low momentum part of the distribution. From the angular distribution,
we see that the pion production off the nucleon model of
Ref.~\cite{Hernandez:2007qq} leads to a forward peaked cross section.
FSI, through quasielastic collisions tends to soften the curve and
leads to a better description of the backward scattering. From our
point of view, these figures might indicate that some mechanism for
pion production, that provides forward high energy pions could be
missing in our theoretical scheme.

\subsection{$NC$ production}
\begin{figure}[tbh]
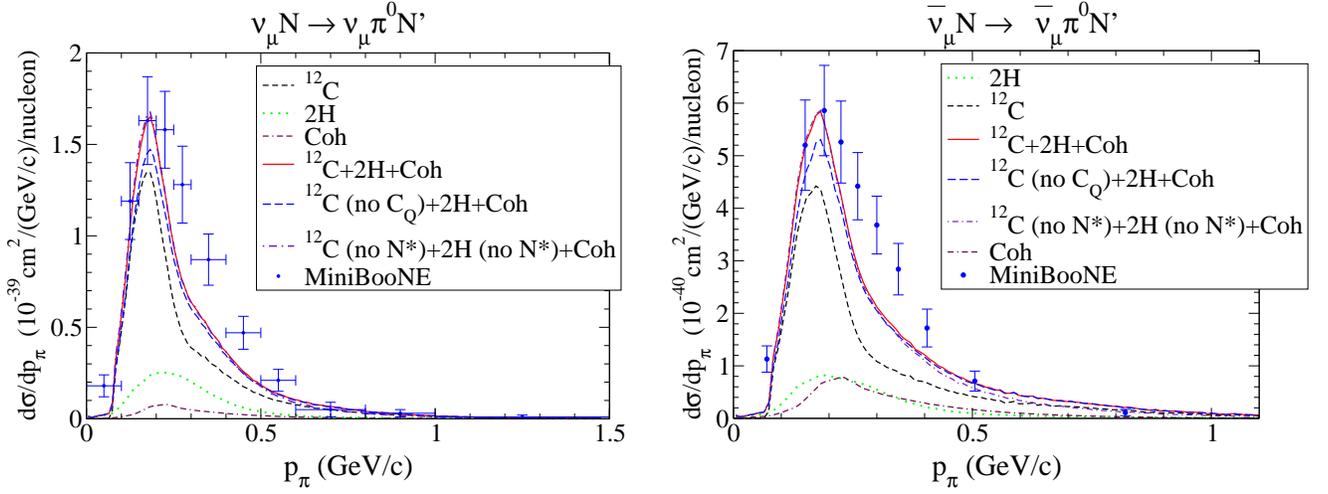

 \center
\includegraphics[height=6.5cm]{3ncpi0mom.eps}\hspace{.5cm}
\includegraphics[height=6.5cm]{4ncpi0momAnu.eps}
\caption{ Flux-folded differential $\frac{d\sigma}{dp_\pi}$ cross
  section per nucleon for $NC$ $1\pi^0$ production by $\nu_\mu$ (left
  panel) and $\bar\nu_\mu$ (right panel) in mineral oil. Short-dashed
  line: $^{12}C$ contribution. Dotted line: $H_2$ contribution.
  Double-dashed dotted line: Coherent contribution. Solid: Full model
  result. Long-dashed line: $^{12}C$ contribution without the $C_Q$
  term (Eq.~(\ref{eq:cq})). Dashed-dotted line:$^{12}C$ contribution without the $D_{13}$
  term.  Data from Ref.~\cite{AguilarArevalo:2009ww}. }
  \label{fig:ncmom}
\end{figure}

In these channels, we compare our model results to data from
Ref.~\cite{AguilarArevalo:2009ww}. In each case we use the different
neutrino/antineutrino fluxes reported by the MiniBooNE
collaboration. In Fig.~\ref{fig:ncmom}, we show the
$\frac{d\sigma}{dp_\pi}$ differential cross section and the different
contributions coming from $^{12}C$, $H_2$ and the coherent production
on $^{12}C$. Both for neutrino and antineutrino reactions, we see that
the model agrees better with data than in the $CC$ case.  We still
obtain cross sections below data in the $0.25\sim0.5\,$GeV/c momentum
region.  We also see that the role of the $D_{13}$ is negligible while
the effects of the $C_Q$ term are  small.

\begin{figure}[tbh]
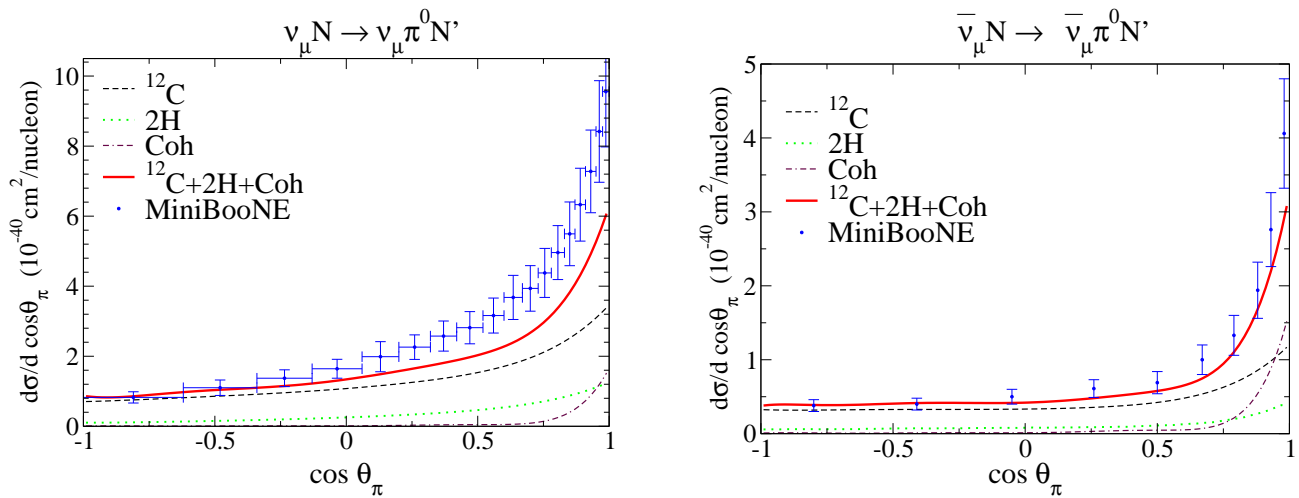

 \center
\includegraphics[height=6.5cm]{3ncpi0cos.eps}\hspace{1cm}
\includegraphics[height=6.5cm]{4ncpi0cosAnu.eps}
\caption{ Flux-folded differential $\frac{d\sigma}{d\cos\theta_\pi}$ 
cross section per nucleon for $NC$ $1\pi^0$ production by $\nu_\mu$ (left panel)
and $\bar\nu_\mu$ (right panel) in mineral
oil. Captions as in Fig.~\ref{fig:ncmom}. Data
 from Ref.~\cite{AguilarArevalo:2009ww}. }
  \label{fig:nccos}
\end{figure}

In Fig.~\ref{fig:nccos} we show now the flux-folded differential
$\frac{d\sigma}{d\cos\theta_\pi}$ cross section. The full model agrees
better with data in the antineutrino case where our results are within
error bars except in the very forward direction. The role and size of
the coherent piece is crucial for the agreement.

There is some deficit for $cos \theta_\pi > 0$ for the reaction with
neutrinos but the agreement, as it was the case for the
$\frac{d\sigma}{dp_\pi}$ differential cross section, is better than in
the corresponding $CC$ reaction. In fact, a minimal enhancement of the
coherent process and/or the hydrogen contribution, that could be
obtained by a larger value of $C_5^A(0)$ also suggested by the CC
results, could lead to a better agreement.

\section{Conclusions}
\label{sec:conclusions}
We have extended the model for pion production by neutrinos on
nucleons from Refs.~\cite{Hernandez:2007qq,Hernandez:2010bx} by the
inclusion of a new term related to the production and decay of the
$D_{13}(1520)$ resonance. The resulting model is expected to give a
fair reproduction of experimental data on nucleons for neutrino
energies up to 2\,GeV, always with the uncertainties associated with
the poor knowledge of some relevant form factors.

 Including nuclear medium corrections that affect both the elementary
 $\pi$ production mechanisms and the pion FSI, we have studied pion
 production by neutrinos in mineral oil in order to compare with
 recent experimental total and differential cross section 
 measurements. The model provides an overall acceptable
 description of data, better for $NC$ than for $CC$ channels. In the
 $CC$ channels,  the predicted total cross sections are
 below data for neutrino energies above $0.8\sim0.9$\,GeV. This result
 is in agreement with other theoretical
 calculations~\cite{Sobczyk.:2012zj,Lalakulich:2012cj}. Differential
 cross sections, folded with the full neutrino flux, show that most of
 the missing pions lie on the forward direction and at high energies.
 This might suggest the need of further production mechanisms.  $NC$
 channels show a better agreement, although theory is also below
 data. We find that the role of the coherent $\pi$ production is
 essential to properly describe the angular distributions in the $NC$
 channels.  We also find that flux unfolded results seem to support a
 large value of $C_5^A(0)$, closer to the PCAC prediction than the
 value obtained from previous combined fits to the ANL and BNL bubble
 chamber data. Actually, as already discussed in
 Ref.~\cite{Lalakulich:2012cj}, it seems that MiniBooNE data would be
 better described using a $C_5^A(0)$ value fitted to BNL data alone.

\begin{acknowledgments}
 This research was supported by  the Spanish Ministerio de Econom\'{\i}a y 
 Competitividad and European FEDER funds
under Contracts Nos.  FIS2011-28853-C02-01, FIS2011-28853-C02-02, FPA2010-
21750-C02-02  and the Spanish Consolider-Ingenio 2010
Programme CPAN (CSD2007-00042), by Generalitat
Valenciana under Contract No. PROMETEO/20090090
and by the EU HadronPhysics3 project, Grant Agreement
No. 283286. 
  \end{acknowledgments}

\appendix

\section{$D_{13}$ resonance contribution to neutrino-nucleon scattering}
\label{app:d13}
In this appendix we give full details of the $D_{13}P$ and $CD_{13}P$ hadronic matrix
elements of the $j^\mu_{cc+}(0)$ weak current, depicted in
Fig.~\ref{fig:d13}. Those we shall call  for short 
$j^\mu_{cc+}|_{DP}$ and $j^\mu_{cc+}|_{CDP}$.

 The  matrix element for the direct ($DP$) contribution is given by
\begin{eqnarray} 
j^\mu_{cc^+}|_{DP}&=&iC^{DP}g_D\sqrt{\frac23}\cos\theta_C
\frac{k_\pi^\alpha}{p_D^2-M_D^2+iM_D\Gamma_D}
\bar u(\vec{p}\,' )\gamma_5 P^D_{\alpha\beta}(p_D) \Gamma^{\beta\mu}_D
\left(p,q \right)
u(\vec{p}\,),\ p_D=p+q,\ C^{DP}=\left\{\begin{array}{ll}
0& p\pi^+\\
1& n\pi^+
\end{array}\right.\nonumber\\
\eea
with $M_D=1520\,$MeV the mass of the $D_{13}$ resonance. For $g_D$ we
take $g_D=20\,$GeV$^{-1}$ which results from 
a fit of the  $D_{13}\to N\pi$ decay width. For the latter we take 
61\% of 115\,MeV. The width has two main contributions, the $N\pi$  and
the $\Delta\pi$ channels. In the propagator we shall use
\beas
\Gamma_D=\Gamma_D^{N\pi}+\Gamma_D^{\Delta\pi},
\eeas  
where for $\Gamma_D^{N\pi}$ we take
\beas
\Gamma_D^{N\pi}=\frac{g_D^2}{8\pi}\frac1{3s}[(\sqrt{s}-M)^2-m_\pi^2]|\vec p_\pi|^3
\theta(\sqrt{s}-M-m_\pi),
\eeas
with $s=p_D^2$ and $|\vec
p_\pi|=\frac{\lambda^{1/2}(s,M^2,m_\pi^2)}{2\sqrt{s}}$,
being $\lambda(a,b,c)=a^2+b^2+c^2-2ab-2ac-2bc$.

For $\Gamma_D^{\Delta\pi}$ we assume an $S-$wave decay and take
\beas
\Gamma_D^{\Delta\pi}=0.39\times115\,{\rm MeV}
\frac{|\vec p_\pi^{\ \prime}|}{|\vec p_\pi^{\ \prime\, o-s}|}\theta(\sqrt{s}-M-m_\pi),
\eeas
with $|\vec
p_\pi^{\ \prime}|=\frac{\lambda^{1/2}(s,M^2_\Delta,m_\pi^2)}{2\sqrt{s}}$ and 
$|\vec
p_\pi^{\ \prime\,o-s}|=\frac{\lambda^{1/2}(M_D^2,M^2_\Delta,m_\pi^2)}{2M_D}$.\\

Besides,
\bea
P^D_{\alpha\beta}(p_D)=-(\slashchar{p}_D+M_D)\bigg[g_{\alpha\beta}
-\frac13\gamma_\alpha\gamma_\beta-\frac23\frac{p_{D\alpha}p_{D\beta}}{M_D^2}
+\frac13\frac{p_{D\alpha}\gamma_\beta-p_{D\beta}\gamma_\alpha}{M_D}\bigg]
\eea
and
\begin{eqnarray}
\Gamma^{\beta\mu} (p,q) &=&\left [ \frac{\tilde C_3^V}{M}\left(g^{\beta\mu}
\slashchar{q}-
q^\beta\gamma^\mu\right) + \frac{\tilde C_4^V}{M^2} \left(g^{\beta\mu}
q\cdot p_D- q^\beta p_D^\mu\right)
+ \frac{\tilde C_5^V}{M^2} \left(g^{\beta\mu}
q\cdot p- q^\beta p^\mu\right)\right. + \tilde C_6^V g^{\beta\mu}
\bigg ] \nonumber\\
&&\hspace*{-.1275cm}+ \left [ \frac{\tilde C_3^A}{M}\left(g^{\beta\mu}
\slashchar{q}-
q^\beta\gamma^\mu\right) + \frac{\tilde C^A_4}{M^2} \left(g^{\beta\mu}
q\cdot p_D- q^\beta p_D^\mu\right)
+ \tilde C_5^A g^{\beta\mu} + \frac{\tilde C_6^A}{M^2} q^\beta q^\mu
\right ]\gamma_5,\ \ p_D=p+q.
\label{eq:d13-2}
\eea
The axial  form factors are taken from Ref.~\cite{Lalakulich:2006sw}
\begin{eqnarray}
\tilde C_3^A=\tilde C_4^A=0,\ \ \tilde C_5^A= \frac{-2.1}{(1-q^2/M_A^2)^2}\frac1{1-q^2/(3M_A^2)},\ \ 
\tilde C_6^A(q^2) = \tilde C_5^A(q^2)\frac{M^2}{m_\pi^2-q^2},\ \ M_A=1\,{\rm
GeV},
\end{eqnarray}
while for the vector ones we fitted  
 the  form factor results in Ref.~\cite{Leitner:2009zz} to get
\begin{eqnarray}
\tilde C_3^V=\frac{-2.98}{[1-q^2/(1.4M_V^2)]^2},\ \ 
\tilde C_4^V=\frac{4.21/D_V}{1-q^2/(3.7M_V^2)},\ \ 
\tilde C_5^V=\frac{-3.13/D_V}{1-q^2/(0.42M_V^2)},\ \ 
\tilde C_6^V=0,
\end{eqnarray}
with $M_V=0.84\,$GeV and $D_V=(1-q^2/M_V^2)^2$.\\

For the  crossed $CDP$ contribution we have
\bea
j^\mu_{cc^+}|_{CDP}&=&-iC^{CDP}g_D\sqrt{\frac23}\cos\theta_C
\frac{k_\pi^\alpha}{p_D^2-M_D^2+iM_D\Gamma_D}
\bar u(\vec{p}\,' )\hat \Gamma^{\mu\beta}_D
\left(p',-q \right) P^D_{\beta\alpha}(p_D) 
\gamma_5 u(\vec{p}\,)\
\eea 
with 
\bea
p_D=p'-q\ \ ,\ \  C^{CDP}=\left\{\begin{array}{ll}
1& p\pi^+\\
0& n\pi^+
\end{array}\right.
\eea
and
\beas
\hat \Gamma^{\mu\beta}_D
\left(p',-q \right)=\gamma^0[\Gamma^{\beta\mu}_D
\left(p',-q \right)]^\dag\gamma^0.
\eeas

From isospin symmetry we further have~\cite{Hernandez:2007qq}
\beas
\langle p\pi^0|j^\mu_{cc^+}(0)|n\rangle&=&-\frac1{\sqrt2}\Big[
\langle p\pi^+|j^\mu_{cc^+}(0)|p\rangle-\langle n\pi^+|j^\mu_{cc^+}|n\rangle
\Big],\\
\langle p\pi^-|j^\mu_{cc^-}(0)|p\rangle&=&\langle n\pi^+|j^\mu_{cc^+}|n\rangle,\\
\langle n\pi^-|j^\mu_{cc^-}(0)|n\rangle&=&\langle p\pi^+|j^\mu_{cc^+}|p\rangle.
\eeas

For $NC$ neutrino and antineutrino induced reactions the contribution 
from the
$D_{13}$ has isovector plus isoscalar parts. The isovector (IV) part is related to
the $CC$ processes and is given by
(see discussion on Sec. III
of Ref.~\cite{Hernandez:2007qq})
\beas
\langle p\pi^0|j^\mu_{nc,{\rm IV}}(0)|p\rangle&=&\frac1{\sqrt2\cos\theta_C}\Big\{
\ \ (1-2\sin^2\theta_W)\Big[\langle p\pi^+|V^\mu_{cc^+}(0)|p\rangle+
\langle n\pi^+|V^\mu_{cc^+}(0)|n\rangle\Big]\\
&&\hspace{2cm}-\Big[\langle p\pi^+|A^\mu_{cc^+}(0)|p\rangle+
\langle n\pi^+|A^\mu_{cc^+}(0)|n\rangle\Big]\Big\},\\
\langle n\pi^+|j^\mu_{nc,{\rm IV}}(0)|p\rangle&=&-\frac1{\cos\theta_C}\Big\{
\ \ (1-2\sin^2\theta_W)\Big[\langle p\pi^+|V^\mu_{cc^+}(0)|p\rangle-
\langle n\pi^+|V^\mu_{cc^+}(0)|n\rangle\Big]\\
&&\hspace{2cm}-\Big[\langle p\pi^+|A^\mu_{cc^+}(0)|p\rangle-
\langle n\pi^+|A^\mu_{cc^+}(0)|n\rangle\Big]\Big\},\\
\langle n\pi^0|j^\mu_{nc,{\rm IV}}(0)|n\rangle&=&\langle p\pi^0|j^\mu_{nc,{\rm
IV}}(0)|p\rangle,\\
\langle p\pi^-|j^\mu_{nc,{\rm IV}}(0)|n\rangle&=&-\langle n\pi^+|j^\mu_{nc,{\rm
IV}}(0)|p\rangle.
\eeas
where we have written 
$j^\mu_{cc^+}(0)=V^\mu_{cc^+}(0)-A^\mu_{cc^+}(0)$ being $V^\mu_{cc^+}(0)$ and
$A^\mu_{cc^+}(0)$ respectively the vector and axial part of the current.

The isoscalar current is given in terms of the isoscalar part of the
electromagnetic current as 
\beas
j^\mu_{nc,{\rm IS}}(0)=-4\sin^2\theta_Ws^\mu_{{\rm em},{\rm IS}}(0),
\eeas
and one has (see discussion on Sec. III
of Ref.~\cite{Hernandez:2007qq})
\beas
\langle n\pi^+|s^\mu_{{\rm em},{\rm IS}}(0)|p\rangle=
\langle p\pi^-|s^\mu_{{\rm em},{\rm IS}}(0)|n\rangle=\sqrt2\,
\langle p\pi^0|s^\mu_{{\rm em},{\rm IS}}(0)|p\rangle=
-\sqrt2\,
\langle n\pi^0|s^\mu_{{\rm em},{\rm IS}}(0)|n\rangle.
\eeas
The direct  contribution to 
$\frac12[ \langle n\pi^0|s^\mu_{{\rm em},{\rm IS}}(0)|n\rangle-\langle p\pi^0|s^\mu_{{\rm em},{\rm IS}}(0)|p\rangle ]$ 
is given by
\beas 
-ig_D\frac1{\sqrt{3}}
\frac{k_\pi^\alpha}{p_D^2-M_D^2+iM_D\Gamma_D}
\bar u(\vec{p}\,' )\gamma_5 P^D_{\alpha\beta}(p_D) \Gamma^{V,{\rm IS}\,\beta\mu}_D
\left(p,q \right)
u(\vec{p}\,),\ p_D=p+q,
\eeas
and
\beas
\Gamma^{V,{\rm IS}\,\beta\mu}_D=\left[ 
\frac{\tilde C_3^{V,{\rm IS}}}{M}\left(g^{\beta\mu}
\slashchar{q}-
q^\beta\gamma^\mu\right) + \frac{\tilde C_4^{V,{\rm IS}}}{M^2} \left(g^{\beta\mu}
q\cdot p_D- q^\beta p_D^\mu\right)
+ \frac{\tilde C_5^{V,{\rm IS}}}{M^2} \left(g^{\beta\mu}
q\cdot p- q^\beta p^\mu\right)\right. + \tilde C_6^{V,{\rm IS}} g^{\beta\mu}
\bigg]
\eeas
with
\beas
\tilde C_3^{V,{\rm IS}}=\frac{-1.21}{[1-q^2/(1.4M_V^2)]^2},\ \ 
\tilde C_4^{V,{\rm IS}}=\frac{0.515/D_V}{1-q^2/(3.7M_V^2)},\ \ 
\tilde C_5^{V,{\rm IS}}=\frac{0.395/D_V}{1-q^2/(0.42M_V^2)},\ \ 
\tilde C_6^{V,{\rm IS}}=0,
\eeas
with their values at $q^2=0$ extracted from Ref.~\cite{Leitner:2009zz}.

The crossed  contribution to 
$\frac12[ \langle n\pi^0|s^\mu_{{\rm em},{\rm IS}}(0)|n\rangle-\langle p\pi^0|s^\mu_{{\rm em},{\rm IS}}(0)|p\rangle ]$ 
is given by
\beas
ig_D\frac1{\sqrt{3}}
\frac{k_\pi^\alpha}{p_D^2-M_D^2+iM_D\Gamma_D}
\bar u(\vec{p}\,' )\hat \Gamma^{V,{\rm IS}\,\mu\beta}_D
\left(p',-q \right) P^D_{\beta\alpha}(p_D) 
\gamma_5 u(\vec{p}\,),\ p_D=p'-q,
\eeas
with
\beas
\hat \Gamma^{V,{\rm IS}\,\mu\beta}_D
\left(p',-q \right)=\gamma^0[\Gamma^{V,{\rm IS}\,\beta\mu}_D
\left(p',-q \right)]^\dag\gamma^0.
\eeas

\end{document}